\documentstyle[12pt]{article}

\catcode`\@=11

\def\psfortextures{
\def\PSspeci@l##1##2{%
\special{illustration ##1\space scaled ##2}}}

\def\psfordvips{
\def\PSspeci@l##1##2{%
\d@my=0.1bp \d@mx=\drawingwd \divide\d@mx by\d@my%
\includegraphics{##1\space}}}

\def\psforoztex{
\def\PSspeci@l##1##2{%
\special{##1 \space
      ##2 1000 div dup scale
      \putsp@ce{\number-\psllx} \putsp@ce{\number-\pslly} translate}}}
\def\putsp@ce#1{#1 }

\def\psonlyboxes{
\def\PSspeci@l##1##2{%
\at{0cm}{0cm}{\boxit{\vbox to\drawinght
  {\vss
  \hbox to\drawingwd{\at{0cm}{0cm}{\hbox{(##1)}}\hss}
  }}}}}

\newdimen\drawinght\newdimen\drawingwd
\newdimen\psxoffset\newdimen\psyoffset
\newbox\drawingBox

\newread\epsffilein    
\newif\ifepsffileok    
\newif\ifepsfbbfound   
\newif\ifepsfverbose   
\newdimen\epsfxsize    
\newdimen\epsfysize    
\newdimen\epsftsize    
\newdimen\epsfrsize    
\newdimen\epsftmp      
\newdimen\pspoints     
\def\ReadPSize#1{
\edef\PSfilename{#1}
\global\def\epsfllx{72}
\global\def\epsflly{72}
\global\def\epsfurx{540}
\global\def\epsfury{720}
\openin\epsffilein=#1
\ifeof\epsffilein\errmessage{I couldn't open #1, will ignore it}\else
   {\epsffileoktrue \chardef\other=12
    \def\do##1{\catcode`##1=\other}\dospecials \catcode`\ =10
    \loop
       \read\epsffilein to \epsffileline
       \ifeof\epsffilein\epsffileokfalse\else
          \expandafter\epsfaux\epsffileline:. \\%
       \fi
   \ifepsffileok\repeat
   \ifepsfbbfound\else
    \ifepsfverbose\message{No bounding box comment in #1; 
using defaults}\fi\fi
   }\closein\epsffilein\fi
\def\psllx{\epsfllx}\def\pslly{\epsflly}%
\def\psurx{\epsfurx}\def\psury{\epsfury}%
\drawinght=\epsfury bp%
\advance\drawinght by-\epsflly bp%
\drawingwd=\epsfurx bp%
\advance\drawingwd by-\epsfllx bp%
}

{\catcode`\%=12 \global\let\epsfpercent=
\long\def\epsfaux#1#2:#3\\{\ifx#1\epsfpercent
   \def\testit{#2}\ifx\testit\epsfbblit
      \epsfgrab #3 @ @ @ \\%
      \epsffileokfalse
      \global\epsfbbfoundtrue
   \fi\else\ifx#1\par\else\epsffileokfalse\fi\fi}%

\def\epsfgrab #1 #2 #3 #4 #5\\{%
   \global\def\epsfllx{#1}\ifx\epsfllx\empty
      \epsfgrab #2 #3 #4 #5 @\\\else
   \global\def\epsflly{#2}%
   \global\def\epsfurx{#3}\global\def\epsfury{#4}\fi}%

\newcount\xscale \newcount\yscale \newdimen\pscm\pscm=1cm
\newdimen\d@mx \newdimen\d@my
\let\ps@nnotation=\relax
\def\psboxto(#1;#2)#3{\vbox{
   \catcode`\:=12
   \ReadPSize{#3}
   \divide\drawingwd by 1000
   \divide\drawinght by 1000
   \d@mx=#1
   \ifdim\d@mx=0pt\xscale=1000
         \else \xscale=\d@mx \divide \xscale by \drawingwd\fi
   \d@my=#2
   \ifdim\d@my=0pt\yscale=1000
         \else \yscale=\d@my \divide \yscale by \drawinght\fi
   \ifnum\yscale=1000
         \else\ifnum\xscale=1000\xscale=\yscale
                    \else\ifnum\yscale<\xscale\xscale=\yscale\fi
              \fi
   \fi
   \divide \psxoffset by 1000\multiply\psxoffset by \xscale
   \divide \psyoffset by 1000\multiply\psyoffset by \xscale
   \global\divide\pscm by 1000
   \global\multiply\pscm by\xscale
   \multiply\drawingwd by\xscale \multiply\drawinght by\xscale
   \ifdim\d@mx=0pt\d@mx=\drawingwd\fi
   \ifdim\d@my=0pt\d@my=\drawinght\fi
\message{[#3\space [BoundingBox\string:
\space\epsfllx\space\epsflly\space\epsfurx\space\epsfury]}%
\message{[scaled\space\the\xscale\string:
\space\the\drawingwd\space x \the\drawinght]]}%
 \hbox to\d@mx{\hss\vbox to\d@my{\vss
   \global\setbox\drawingBox=\hbox to 0pt{\kern\psxoffset\vbox to 0pt{
      \kern-\psyoffset
      \PSspeci@l{\PSfilename}{\the\xscale}
      \vss}\hss\ps@nnotation}
   \global\ht\drawingBox=\the\drawinght
   \global\wd\drawingBox=\the\drawingwd
   \baselineskip=0pt
   \copy\drawingBox
 \vss}\hss}
  \global\psxoffset=0pt
  \global\psyoffset=0pt
  \global\pscm=1cm
  \global\drawingwd=\drawingwd
  \global\drawinght=\drawinght
}}

\def\psboxscaled#1#2{\vbox{
  \catcode`\:=12
  \ReadPSize{#2}
  \xscale=#1
  \divide\drawingwd by 1000\multiply\drawingwd by\xscale
  \divide\drawinght by 1000\multiply\drawinght by\xscale
  \divide \psxoffset by 1000\multiply\psxoffset by \xscale
  \divide \psyoffset by 1000\multiply\psyoffset by \xscale
  \global\divide\pscm by 1000
  \global\multiply\pscm by\xscale
\message{[#2\space [BoundingBox\string:
\space\epsfllx\space\epsflly\space\epsfurx\space\epsfury]}%
\message{[scaled\space\the\xscale\string:
\space\the\drawingwd\space x \the\drawinght]]}%
  \global\setbox\drawingBox=\hbox to 0pt{\kern\psxoffset\vbox to 0pt{
     \kern-\psyoffset
     \PSspeci@l{\PSfilename}{\the\xscale}
     \vss}\hss\ps@nnotation}
  \global\ht\drawingBox=\the\drawinght
  \global\wd\drawingBox=\the\drawingwd
  \baselineskip=0pt
  \copy\drawingBox
  \global\psxoffset=0pt
  \global\psyoffset=0pt
  \global\pscm=1cm
  \global\drawingwd=\drawingwd
  \global\drawinght=\drawinght
}}

\def\psannotate#1#2{\def\ps@nnotation{#2\global\let\ps@nnotation=\relax}#1}
\def\pscaption#1#2{\vbox{
   \setbox\drawingBox=#1
   \copy\drawingBox
   \vskip\baselineskip
   \vbox{\hsize=\wd\drawingBox\setbox0=\hbox{#2}
     \ifdim\wd0>\hsize
       \noindent\unhbox0\tolerance=5000
    \else\centerline{\box0}
    \fi
}}}

\def\at#1#2#3{\setbox0=\hbox{#3}\ht0=0pt\dp0=0pt
  \rlap{\kern#1\vbox to0pt{\kern-#2\box0\vss}}}

\newdimen\gridht \newdimen\gridwd
\def\gridfill(#1;#2){
  \setbox0=\hbox to 1\pscm
  {\vrule height1\pscm width.4pt\leaders\hrule\hfill}
  \gridht=#1
  \divide\gridht by \ht0
  \multiply\gridht by \ht0
  \advance \gridht by \ht0
  \gridwd=#2
  \divide\gridwd by \wd0
  \multiply\gridwd by \wd0
  \advance \gridwd by \wd0
  \vbox to \gridht{\leaders\hbox to\gridwd{\leaders\box0\hfill}\vfill}}

\def\frameit#1#2#3{\hbox{\vrule width#1\vbox{
  \hrule height#1\vskip#2\hbox{\hskip#2\vbox{#3}\hskip#2}%
        \vskip#2\hrule height#1}\vrule width#1}}
\def\boxit#1{\frameit{0.4pt}{0pt}{#1}}

\catcode`\@=12 

\psfordvips

\jot = 1.5ex

\parskip 5pt plus 1pt

\catcode`\@=11

\@addtoreset{equation}{section}
\def\theequation{\thesection\arabic{equation}}

\def\@normalsize{\@setsize\normalsize{15pt}\xiipt\@xiipt
\abovedisplayskip 14pt plus3pt minus3pt%
\belowdisplayskip \abovedisplayskip
\abovedisplayshortskip  \z@ plus3pt%
\belowdisplayshortskip  7pt plus3.5pt minus0pt}
\def\small{\@setsize\small{13.6pt}\xipt\@xipt
\abovedisplayskip 13pt plus3pt minus3pt%
\belowdisplayskip \abovedisplayskip
\abovedisplayshortskip  \z@ plus3pt%
\belowdisplayshortskip  7pt plus3.5pt minus0pt
\def\@listi{\parsep 4.5pt plus 2pt minus 1pt
            \itemsep \parsep
            \topsep 9pt plus 3pt minus 3pt}}

\def\underline#1{\relax\ifmmode\@@underline#1\else
        $\@@underline{\hbox{#1}}$\relax\fi}
\@twosidetrue
\relax

\catcode`@=12

\evensidemargin 0.0in
\oddsidemargin 0.0in
\topmargin -0.2in
\textwidth 6.4in
\textheight 8.9in

\catcode`\@=11

\def\section{\@startsection{section}{1}{\z@}{3.5ex plus 1ex minus
   .2ex}{2.3ex plus .2ex}{\large\bf}}
\def\thesection{\arabic{section}.}

\def\ps@headings{\def\@oddfoot{}\def\@evenfoot{}
\def\@oddhead{\hbox{}\hfill
        \makebox[.5\textwidth]{\raggedright\ignorespaces --\thepage{}--
        \hfill }}
\def\@evenhead{\@oddhead}
\def\subsectionmark##1{\markboth{##1}{}}
}

\ps@headings

\catcode`\@=12

\relax

\def\figcap{\section*{Figure Captions\markboth
        {FIGURECAPTIONS}{FIGURECAPTIONS}}\list
        {Fig. \arabic{enumi}:\hfill}{\settowidth\labelwidth{Fig. 999:}
        \leftmargin\labelwidth
        \advance\leftmargin\labelsep\usecounter{enumi}}}
 \relax
\def\tablecap{\section*{Table Captions\markboth
        {TABLECAPTIONS}{TABLECAPTIONS}}\list
        {Table \arabic{enumi}:\hfill}{\settowidth\labelwidth{Table 999:}
        \leftmargin\labelwidth
        \advance\leftmargin\labelsep\usecounter{enumi}}}
 \relax
\def\reflist{\section*{References\markboth
        {REFLIST}{REFLIST}}\list
        {[\arabic{enumi}]\hfill}{\settowidth\labelwidth{[999]}
        \leftmargin\labelwidth
        \advance\leftmargin\labelsep\usecounter{enumi}}}
 \relax

\catcode`\@=11

\def\ps@headings{\def\@oddfoot{}\def\@evenfoot{}
\def\@oddhead{\hbox{}\hfill
        \makebox[.5\textwidth]{\raggedright\ignorespaces --\thepage{}--
        \hfill }}
\def\@evenhead{\@oddhead}
\def\subsectionmark##1{\markboth{##1}{}}
}

\ps@headings

\relax

\def\firstpage#1#2#3#4#5#6{
\begin{document}
\begin{titlepage}
\nopagebreak
\title{\begin{flushright}
        \vspace*{-1.8in}
        {\normalsize CERN--TH/96-211}\\[-9mm]
   {\normalsize CPTH--S462.0796}\\[-9mm]
        {\normalsize hep-th/9608012}\\[4mm]
\end{flushright}
\vfill
{#3}}
\author{\large #4 \\[1cm] #5}
\maketitle
\vskip -7mm     
\nopagebreak 
\begin{abstract}
{\noindent #6}
\end{abstract}
\vfill
\begin{flushleft}
\rule{16.1cm}{0.2mm}\\[-3mm]
$^{\star}${\small Research supported in part by\vspace{-4mm}
the National Science Foundation under grant
PHY--93--06906, \linebreak in part by the EEC contract
CHRX-CT93-0340 and in part by
CNRS--NSF grant INT--92--16146.}\\[-3mm] 
$^{\dagger}${\small Laboratoire Propre du CNRS UPR A.0014.}\\[2mm]
CERN--TH/96-211\\[-2mm]
July 1996
\end{flushleft}
\thispagestyle{empty}
\end{titlepage}}
\newcommand{\dal}{\raisebox{0.085cm}
{\fbox{\rule{0cm}{0.07cm}\,}}}
\newcommand{\dt}{\partial_{\langle T\rangle}}
\newcommand{\dtbar}{\partial_{\langle\bar{T}\rangle}}
\newcommand{\al}{\alpha^{\prime}}
\newcommand{\mst}{M_{\scriptscriptstyle \!S}}
\newcommand{\mpl}{M_{\scriptscriptstyle \!P}}
\newcommand{\dv}{\int{\rm d}^4x\sqrt{g}}
\newcommand{\lv}{\left\langle}
\newcommand{\rv}{\right\rangle}
\newcommand{\ph}{\varphi}
\newcommand{\psibar}{\tilde{\psi}}
\newcommand{\Psibar}{\widetilde{\Psi}}
\newcommand{\sbar}{\,\bar{\! S}}
\newcommand{\xbar}{\,\bar{\! X}}
\newcommand{\fbar}{\,\bar{\! F}}
\newcommand{\zbar}{\bar{z}}
\newcommand{\wbar}{\bar{w}}
\newcommand{\dbar}{\,\bar{\!\partial}}
\newcommand{\tbar}{\bar{T}}
\newcommand{\taubar}{\bar{\tau}}
\newcommand{\ubar}{\bar{U}}
\newcommand{\ybar}{\bar{Y}}
\newcommand{\phb}{\bar{\varphi}}
\newcommand{\cm}{Commun.\ Math.\ Phys.~}
\newcommand{\pr}{Phys.\ Rev.\ D~}
\newcommand{\pl}{Phys.\ Lett.\ B~}
\newcommand{\ibar}{\bar{\imath}}
\newcommand{\jbar}{\bar{\jmath}}
\newcommand{\np}{Nucl.\ Phys.\ B~}
\newcommand{\e}{{\rm e}}
\newcommand{\be}{\begin{equation}}
\newcommand{\ee}{\end{equation}}
\newcommand{\gsi}{\,\raisebox{-0.13cm}{$\stackrel{\textstyle
>}{\textstyle\sim}$}\,}
\newcommand{\lsi}{\,\raisebox{-0.13cm}{$\stackrel{\textstyle
<}{\textstyle\sim}$}\,}
\date{}
\firstpage{3118}{IC/95/34}
{\large\bf Aspects of Type I -- Type II -- Heterotic Triality\\[-2mm]
in Four Dimensions$^{\star}$} {I. Antoniadis$^{\,a,b}$, C. Bachas$^{\,a}$, C.
{}Fabre$^{\,a}$, H. Partouche$^{\,a}$, T.R. Taylor$^{\,b,c}$}  
{\normalsize\sl
$^a$Centre de Physique Th\'eorique, Ecole Polytechnique,$^\dagger$
{}F-91128 Palaiseau, France\\[-3mm]
\normalsize\sl $^b$Theory Division, CERN, 1211 Geneva 23, Switzerland\\[-3mm]
\normalsize\sl $^c$Department of Physics, Northeastern
University, Boston, MA 02115, U.S.A.}
{We discuss the equivalence between Type I, Type II and Heterotic $N=2$
superstring theories in four dimensions. We study the
effective field theory of Type I models obtained by orientifold
reductions of Type IIB compactifications on $K_3\times T^2$.
 We show that the
perturbative prepotential is determined
by the one-loop corrections to the Planck mass and is associated
to an index.
As is the case for threshold corrections to gauge couplings, 
 this
renormalization is entirely due to  $N=2$ BPS states that 
originate from $D=6$ massless string modes.
  We apply our  result to the so-called $S$-$T$-$U$ model
which admits simultaneous Type II and Heterotic descriptions,
 and show that all
three prepotentials agree in the appropriate limits as expected from 
the superstring triality conjecture.}
\section{Introduction}

There has been some very convincing evidence accumulated so far
for the equivalence of theories which were believed in the past
to describe truly different types of superstrings. 
Type I, Type II and Heterotic theories seem merely to provide 
complementary  descriptions of a more complicated
theory of fundamental interactions, and the larger framework of superstring 
dualities now includes also M-theory and F-theory descriptions.
In order to reach various points on the web of connected models,
it is often convenient to start from ten dimensions and to descend to 
lower dimensions by compactifying these well-known theories.
The equivalence of various superstring compactifications can then be
understood as a consequence of a few fundamental
dualities originating from higher dimensions \cite{rev}.

Among the four-dimensional models, the most familiar examples of
dual pairs are based on Type II and Heterotic
constructions \cite{kv} whose equivalence originates from
the well-established six-dimensional duality between 
Type IIA compactified on $K_3$ and Heterotic compactified on $T^4$ \cite{ht}.
These models have $N=4$ or $N=2$ spacetime supersymmetry in $D=4$ and their
equivalence has been checked in many ways, including some highly non-trivial
quantitative comparisons of the  respective low-energy effective actions
\cite{kv,checks,klm}. As a generic feature,
 the string coupling of the Heterotic
side is mapped under such duality to a ``geometric'' modulus on the Type II
side.

Type I theory remained a wild card in duality conjectures until
quite recently Polchinski and Witten presented several
arguments for the equivalence of Type I and Heterotic
theories in ten dimensions \cite{pw}. Although in $D=10$ this is 
a strong-weak coupling duality, it turns out that upon appropriate
compactification to $D=4$ one obtains $N=2$ supersymmetric
dual pairs with the Heterotic gauge coupling
mapped to a Type I gauge coupling in a way that
some weakly coupled regions overlap on both sides. 
This work is focused on Type I -- Heterotic duality in 
$D=4$. We discuss the mapping of special coordinates
of the special K\"ahler manifold describing 
the massless vector multiplet sector. We compute the perturbative prepotential
on the Type I side for a general class of $K_3\times T^2$ orientifold
reductions of the underlying Type IIB theory. We obtain a general expression
which involves $N=2$ BPS states only. It agrees with the appropriate limit of
the corresponding expression in the Heterotic theory. We apply this result 
to a 
specific example, which admits 
all three Type I, Type II and Heterotic descriptions. The agreement of
all three prepotentials provides here a convincing evidence for
a true superstring triality in $D=4$.

The paper is organized as follows. 
In section 2, we review Type I -- Heterotic duality in $D=6$ 
\cite{sw,blpssw,gj}
and recall some basic features of the effective field theory
describing Type I orientifold reductions of Type IIB theory \cite{s1,bs,sw}
that are dual to Heterotic $K_3$ compactifications.
In section 3, we discuss the tree-level effective actions 
of four-dimensional models obtained by toroidal compactifications of 
six-dimensional Type I models. We identify the universal vector
moduli $S$, $T$ and $U$ (in Heterotic notation) and describe
the duality mapping of special coordinates. In section 4 we discuss 
quantum corrections, explaining what kind of useful information can 
be extracted from purely perturbative Type I computations.
We discuss the problem of determination of the one-loop prepotential. In the
Heterotic theory one can extract it from the universal part
of threshold corrections to gauge couplings \cite{agnt,prep,hm}. In Type I
theory, the one-loop threshold corrections have recently been analyzed
in
ref.\cite{bf}; however it is not possible to extract from them the K\"ahler
metric. The reason is that unlike the Heterotic case, the Planck mass
receives non-vanishing corrections which force redefinitions of 
special coordinates at the one-loop level. As a result,
the universal part of threshold corrections is absorbed into the tree-level
gauge coupling. However the K\"ahler metric, hence also the perturbative
prepotential, can be extracted from the one-loop 
Planck mass. In section 5, we present the one-loop computation of the Planck 
mass
which allows a determination of the K\"ahler metric in Type I theory.
The metric, hence also the prepotential are completely determined by the BPS
spectrum of the theory. Some technical details  and the expressions for
various propagators used in the computation are given in the Appendix.
In section 6, we consider a specific orientifold model with the
massless spectrum consisting of 3 vector multiplets and 244 hypermultiplets
\cite{blpssw,gj}. This is the so-called $S$-$T$-$U$ model which admits
simultaneous Type II and Heterotic descriptions \cite{kv};
 the exact form of its
prepotential has been determined before by using Type II -- Heterotic duality
\cite{klm}. We apply the general formula of section 5 to determine the one-loop
K\"ahler metric. The result reproduces the $T\rightarrow i\infty$ limit of the
Heterotic case
\cite{prep}, as expected from duality. We summarize our results in section 7.

\section{Type I Effective Field Theory in Six Dimensions}

In this work, we will consider four-dimensional $N=2$ Type I superstrings
obtained by toroidal compactifications of $D=6$, $N=1$ models.
 The latter can be
constructed as orientifold compactifications of Type IIB theories
\cite{bs,gp}. In this section we review some basic features of the effective
field theory in $D=6$, in connection with Type I -
 Heterotic duality \cite{sw}.

Anomaly cancellation constrains the massless spectrum to satisfy 
\begin{equation}
n_H-n_V=244-29(n_T-1)\ , \label{ac}
\end{equation}
where $n_H$, $n_V$ and $n_T$ are the numbers of hyper, vector and tensor
multiplets, respectively. Since we are interested in theories dual
to Heterotic compactifications, we restrict our discussion to $n_T=1$.
These theories contain one two-index antisymmetric tensor field whose
self-dual part belongs to the tensor multiplet while its anti-self-dual
part belongs to the gravitational multiplet. In Type I theory the
antisymmetric tensor arises from the Ramond-Ramond (R-R) sector. 

The scalar component of the tensor multiplet is related to the
$K_3$ volume and determines the gauge coupling constants. In fact,
a standard dimensional reduction from $D=10$ to $D=6$ gives
\begin{equation}
{\cal L}^{(6)}= -e^{-2\phi_6}\left\{\frac{1}{2}R^{(6)}+
2\left(\frac{\partial\omega}{\omega}\right)^2-2(\partial\phi_6)^2\right\}
-\frac{1}{4}e^{-\phi_6}\omega^2F^2
-\frac{1}{16}\omega^4(dB-\Omega)^2+\dots      \label{bos6}
\end{equation}
Here, $\omega^4$ is the volume of $K_3$,
$R^{(6)}$ is the scalar curvature, $\phi_6$
is the six-dimensional dilaton, $F$ is the gauge field strength,
$dB$ is the antisymmetric tensor field strength and $\Omega$
is the gauge Chern-Simons term.
It is now easy to see that in the Einstein frame the factor
$e^{-\phi_6}$ drops from the gauge kinetic terms, so that the gauge coupling
constant becomes $1/\omega$. The scalar $\omega$ belongs to the
tensor multiplet while the string coupling $e^{\phi_6}$ belongs to a 
hypermultiplet.

The gauge couplings of eq.(\ref{bos6}), obtained by a simple dimensional 
reduction, are not the most general ones. First of all, in orientifold
constructions, there appear additional gauge bosons associated to
open strings with end-points fixed on 5-branes \cite{gp}. The corresponding
gauge kinetic terms are 
\be
-\frac{1}{4}e^{-\phi_6}\omega^{-2}{F'}^{2}   
-\frac{1}{16}B\wedge F'\wedge F'+\dots  \label{gau6}
\end{equation}
By going again to the Einstein frame, one sees that the coupling
constant becomes $\omega$ for these (primed) gauge fields.
In the most general case the gauge couplings are given by
linear combinations \cite{s1}: 
\be
\frac{1}{g_i^2}=v_i\omega^2+v'_i\omega^{-2} \label{gen}
\ee
where $v_i$ and $v'_i$ are constants. In models involving
non-vanishing constants of both types, {\em i.e}.\ $v_i v_j'\neq 0$ for
some gauge group generators $i, j$, 
an additional complication arises because the effective field theory 
action given by the sum of
eqs.\ (\ref{bos6}) and (\ref{gau6}) is not consistent with
supersymmetry \cite{fms}. The last terms of
 eqs.\ (\ref{bos6}) and (\ref{gau6})
are Wess-Zumino terms which cancel gauge anomalies in analogy
with the Green-Schwarz mechanism in $D=10$.
We will come back to this problem later, after compactifying to $D=4$.

Type I -- Heterotic duality originates from $D=10$, where both
theories are believed to be equivalent after inverting the
string couplings and rescaling the Regge slope $\alpha'$ \cite{pw}:
\be
\phi_{10}^I=-\phi_{10}^H\qquad\qquad \alpha'_I=e^{\phi_{10}^H}\alpha'_H
\label{d10}\ee
where $I$ and $H$ refer to Type I and Heterotic, respectively.
Using the relation between six- and
ten-dimensional dilatons, $e^{-2\phi_6}=e^{-2\phi_{10}}\omega^4$, one finds
\be
e^{-2\phi_{6}^I}=\omega^4_H\qquad\qquad \omega^4_I=e^{-2\phi_{6}^H}
\label{d6}\ee
which implies that the six-dimensional theories become equivalent
after interchanging the square of the string coupling with the inverse of
the $K_3$ volume, $e^{2\phi_{6}}\leftrightarrow\omega^{-4}$.
As a result, the interactions of tensor
and gauge multiplets are purely classical on the Type I side, since the string 
coupling belongs to a hypermultiplet. On the other hand, 
the hypermultiplet sector of the Heterotic side does not receive any
quantum corrections since the string coupling there belongs
to a tensor multiplet. 

\section{Tree-Level Effective Field Theory in Four Dimensions}

After toroidal compactification to $D=4$ one obtains $N=2$ superstring
models with the massless spectrum consisting of
the supergravity multiplet, $n_V+3$ vector multiplets and $n_H$
hypermultiplets. The three additional vector bosons
and the graviphoton arise from the metric and from the
R-R antisymmetric tensor. The way scalar particles
fit into supermultiplets is more subtle, therefore we discuss
them now in some detail.

In addition to the scalar components of the $n_V$ vector multiplets
which are open string states, and the scalar of the tensor multiplet,
there are also 5 scalars which appear upon compactification to $D=4$.
Three of them come from the torus metric $G_{IJ}$ and two from
the antisymmetric tensor: the axion dual to $D=4$
components $B_{\mu\nu}$, and the internal component $B_{IJ}$.
Note that the usual NS-NS (Neveu-Schwarz) antisymmetric tensor is eliminated
by the orientifold projection. Straightforward dimensional reduction
of the $D=6$ effective action (\ref{bos6}), (\ref{gau6}) yields
\begin{eqnarray}
{\cal L}^{(4)}&=&-e^{-2\phi_4}\left\{\frac{1}{2}R^{(4)}+
2\left(\frac{\partial\omega}{\omega}\right)^2-2(\partial\phi_4)^2
-\frac{\partial U\partial \ubar}{(U-\ubar)^2}+
\frac{1}{4}
\left(\frac{\partial\sqrt{G}}{\sqrt{G}}\right)^2\right\}\nonumber\\
& &-\frac{1}{4}e^{-\phi_4}G^{1/4}\omega^2F^2
-\frac{1}{4}e^{-\phi_4}G^{1/4}\omega^{-2}F'^2+\dots
\label{bos4}
\end{eqnarray}
where for the moment we kept only the terms which are relevant for the 
identification of the supermultiplets. The four-dimensional dilaton
(defined as the string coupling constant in $D=4$)
is given by $e^{-2\phi_4}=e^{-2\phi_6}\sqrt{G}$. In eq.(\ref{bos4}),
$U$ is the usual complex modulus which determines the complex structure
of the torus: $U=(G_{45}+i\sqrt{G})/G_{44}$.

Guided by the form of gauge coupling constants, we define two complex
fields $S$ and $S'$, with the imaginary parts given by
\be
S_2=e^{-\phi_4}G^{1/4}\omega^2\qquad\qquad
 S'_2=e^{-\phi_4}G^{1/4}\omega^{-2}\ .
\label{ss}\ee
The real parts $S_1$ and $S'_1$ are defined as the scalar dual to $B_{\mu\nu}$ 
and $B_{45}$, respectively.
In terms of these fields, eq.(\ref{bos4}), transformed into the 
Einstein frame
and supplemented by the dimensionally reduced kinetic term $(dB)^2$ of 
eq.(\ref{bos6}), reads
\begin{eqnarray}
{\cal L}^{(4)}&=&-\frac{1}{2}R^{(4)}
+\frac{\partial U\partial \ubar}{(U-\ubar)^2}
-\frac{(\partial S_1)^2}{4S_2^2}
-\frac{(\partial S'_1)^2}{4{S'}_2^2}
-\frac{(\partial S_2)^2}{4S_2^2}
-\frac{(\partial S'_2)^2}{4{S'}_2^2} \nonumber\\ & &
-\frac{1}{2}(\partial\phi_6)^2
-\frac{1}{4}S_2F^2-\frac{1}{4}S'_2F'^2+\dots
\label{bos}\end{eqnarray}
The complex scalars $S$, $S'$ and $U$ belong to vector multiplets 
while the six-dimensional dilaton $\phi_6$ remains in a hypermultiplet.
Equation (\ref{bos}) shows that in the absence of open string vector 
multiplets,
the three universal scalars $S$, $S'$ and $U$ parameterize a $[SU(1,1)]^3$ 
manifold, with the corresponding prepotential $F=SS'U$. 

Type I theory exhibits two continuous Peccei-Quinn
symmetries associated to $S$ and $S'$ axion shifts which remain
valid to all orders of perturbation theory since
the corresponding axions originate from the R-R sector.
In terms of the independent scalars appearing in eq.(\ref{bos}),
the four-dimensional string coupling is a combination of
fields belonging to hyper and vector multiplets:
\be
e^{-2\phi_4}=e^{-\phi_6}(S_2S'_2)^{1/2}\ .
\label{dil}\ee
This means that both hyper and vector
multiplet sectors can in principle receive quantum corrections 
in four-dimensional Type I theory,
once $e^{-2\phi_4}$ combines with appropriate factors to form
a hyper or a vector multiplet component.

We can now use Type I -- Heterotic relations (\ref{d6})
in $D=6$ to deduce the duality mapping in four dimensions:
\be
S_I=S_H\qquad\qquad S'_I=T_H\quad\qquad U_I=U_H\, . \label{d4}
\ee
Here, $S_H=\alpha +ie^{-2\phi_4^H}$, where $\alpha$ is the axion
dual to $B_{\mu\nu}$ and $\phi_4^H$ is the Heterotic dilaton;
$T_H\equiv T=B_{45}+i\sqrt{G}$ is the usual K\"ahler-class
modulus of the 2-torus. This means that at the exact level
the two theories are equivalent upon identification of $S'$ with $T$
and of the hypermultiplet scalar $e^{-2\phi_6}$ with the $K_3$ volume,
according to eq.(\ref{d6}).

In order to see how the prepotential depends on the additional
$n_V$ open string vector multiplets, we first discuss the
simplest case of vectors obtained by dimensional reduction from
$D=10$. Starting from the Lagrangian (\ref{bos6}) one obtains:
\begin{eqnarray}
{\cal L}^{(4)}&=& -\frac{1}{2}R^{(4)}+
\frac{\partial S\partial \sbar}{(S-\sbar)^2}
+\frac{\partial U\partial \ubar}{(U-\ubar)^2}
-\frac{(\partial S'_2)^2}{4{S'}_2^2}
-\frac{(\partial S'_1+{\textstyle \frac{1}{2}\sum_i}a_4^i\!
\stackrel{\leftrightarrow}{\partial}\! a_5^i)^2}{4{S'}_2^2}\nonumber\\ & &
+~{\textstyle \sum_i}\frac{|U\partial a_4^i-\partial a_5^i|^2}{(S'-\sbar')
(U-\ubar)}+\dots\label{l4}
\end{eqnarray}
where $a_4,a_5$ are the scalars arising from the compact
components of six-dimensional vector fields. These Lagrangian terms
can be derived from the $N=2$ prepotential 
\be
F^{(0)}=S(S'U-{\textstyle\frac{1}{2}\sum_i}A_i^2) \label{pre}
\ee
where the special coordinates of gauge fields are defined by \cite{clm}
\be A_i=a_4^iU-a_5^i\ ,\label{aa}\ee 
and $S'$ is redefined as
\be 
S'=S'|_{A=0}+{\textstyle\frac{1}{2}\sum_i}a_4^iA_i\ .\label{sprime}\ee
The duality transformation
(\ref{d4}) maps $A_i$ into perturbative gauge multiplets on the 
Heterotic side.

Instead of the vectors $A_i$ coming from the ten-dimensional
gauge group [$SO(32)$], consider now the vector multiplets related
to 5-branes discussed in section 2. 
Their  $D=6$ kinetic terms are of the form (\ref{gau6}), and lead
to the same $D=4$ effective Lagrangian and prepotential as in eqs.(\ref{l4})
and (\ref{pre}), with the replacement of $A$ by $A'$ 
[defined as in eq.(\ref{aa})] and with the interchange $S\leftrightarrow S'$. 
On the Heterotic side, these vector multiplets have a non-perturbative origin,
and the corresponding gauge couplings are determined by the $T$
modulus instead of the Heterotic dilaton $S$ (modulo exponentially
suppressed instanton corrections).

A more complicated situation arises in the simultaneous
presence of $A$- and $A'$-type of fields, or 
in the presence of vector multiplets with gauge couplings
involving non-vanishing $v$ and $v'$ as in eq.(\ref{gen}).
If one naively starts from the combined action (\ref{bos6})+(\ref{gau6})
and goes down to $D=4$, one finds that in terms of the redefined
complex fields
\be
S'=S'|_{A=0}+{\textstyle \sum_i\frac{v_i}{2}}a_4^iA_i\qquad\qquad
S=S|_{A=0}+{\textstyle \sum_i\frac{v'_i}{2}}a_4^iA_i\ ,
\label{s1}\ee
the scalar kinetic terms are determined by the K\"ahler potential
\begin{eqnarray}
K &=&-\ln\{(S'-\sbar')(U-\ubar)-{\textstyle\sum_i
\frac{v_i}{2}}(A_i-\bar{A}_i)^2\}
-\ln\{(S-\sbar)(U-\ubar)-{\textstyle\sum_i\frac{v'_i}{2}}(A_i-\bar{A}_i)^2\}
\nonumber\\ & &\qquad +\ln (U-\ubar)\nonumber\\
&=&-\ln\{(S-\sbar)(S'-\sbar')(U-\ubar)
-{\textstyle\frac{1}{2}\sum_i}[v_i(S-\sbar)+v'_i(S'-\sbar')](A_i-\bar{A}_i)^2
\nonumber\\ & &\qquad
+\frac{1}{(U-\ubar)}
[{\textstyle\sum_i\frac{v_i}{2}}(A_i-\bar{A}_i)^2]
[{\textstyle\sum_j\frac{v'_j}{2}}(A_j-\bar{A}_j)^2]\}\ .\label{kahler}
\end{eqnarray}
The corresponding scalar manifold, although K\"ahler, is not
of the special type, which is a consequence of the fact that the
six-dimensional action was not consistent with supersymmetry,
as already mentioned before.
This six-dimensional anomaly disappears in lower dimensions,
where ``anomalous'' terms are canceled by local counterterms.
It is not difficult to realize that the role of the counterterms
is to cancel the last term in eq.(\ref{kahler}),
 so that the K\"ahler manifold
becomes special, as required by $N=2$ supersymmetry.
The corresponding prepotential is
\be
F^{(0)}=SS'U-{\textstyle\frac{1}{2}\sum_i}(v_iS+v'_iS')A_i^2\ .\label{pp}\ee
The above result agrees with the analysis of $D=5$ compactification
of the same theory \cite{fms}. Moreover, it can be verified directly
at the string level in various examples. 

Note that in the case when $v_iv'_i<0$ for some $i$, the corresponding
gauge kinetic term may vanish for finite values of $S$ and $S'$.
This singularity is inherited from the corresponding
term in $D=6$ and is related to the appearance of tensionless
strings \cite{sw}. On the Heterotic side, for perturbative gauge fields, $v$
is the Kac-Moody level while a non-zero $v'$ may arise
from one-loop threshold corrections in the $T\rightarrow i\infty$ limit.

\section{Type I -- Heterotic Duality and Quantum Corrections}

In this section we discuss perturbative
corrections to the prepotential on the Type I side.
We want to understand how duality can be tested by comparing
prepotentials, and eventually what information
can be extracted from purely perturbative Type I computations.
The two Peccei-Quinn symmetries dictate the following form
of Type I prepotential:
\be
F(S,S',U,A)=
 F^{(0)}+f_I(U,A)+\makebox{non-perturbative corrections,}\label{fI}
\ee
where $F^{(0)}$ is the tree-level prepotential (\ref{pp})
and $f_I(U,A)$ is the one-loop correction. 
Type I non-perturbative terms include instanton terms
which are suppressed in the large $S_2$ and/or $S'_2$ limit.
Although $f_I$ cannot depend on $S$ and $S'$ in a continuous way,
its form may be different in the regions $S_2>S'_2$ and $S_2<S'_2$,
{\em i.e}.\ for large and small $K_3$ volumes, {\em c.f}.\ 
eq.(\ref{ss}). In models which are invariant under ``$T$-duality''
($\omega\rightarrow 1/\omega$) one obtains the same result in the
two regions. 

On the Heterotic side, there is only one 
perturbative Peccei-Quinn symmetry (associated to $S$),
therefore the analogous expression is
\be
F(S,T,U,A)= F^{(0)}+f_H(T,U,A)+\makebox{non-perturbative corrections,}
\label{fH}\ee
where we used the duality relation (\ref{d4}) which maps $S'$ into $T$.
Type I -- Heterotic duality implies that
\be
\lim_{T_2\rightarrow\infty}f_H=f_I|_{S_2>S'_2}.\label{lim}\ee
As indicated above, this relation between perturbative prepotentials
is valid only for $S_2>S'_2$,
since in the perturbative expansion of the Heterotic theory 
the large $S$ limit is taken first. The other region,  $S_2<S'_2$,
can only be reached non-perturbatively from the Heterotic side,
therefore Type I perturbation theory can be {\em a priori} useful
in studying the corresponding region $T_2>S_2\rightarrow\infty$.
The two regions can be related, though, by 
$\omega\rightarrow 1/\omega$ duality which corresponds to
non-perturbative $S\leftrightarrow T$ exchange.
Note that if a given model admits also a Type II description,
the full prepotential $F(S,T,U,A)$ can be computed exactly at the 
classical level on the Type II side.

Let us consider now a class of models based on orientifold
reductions of Type IIB theory compactified on the $K_3$ orbifold $T^4/Z_2$
\cite{bs,gp}. In $D=6$ these models have one tensor multiplet and a maximal
gauge group
$U(16)\times U(16)'$. The two group factors
 are associated to open strings with 
Neumann-Neumann (N-N) and Dirichlet-Dirichlet (D-D) boundary conditions,
respectively. In addition, there are massless hypermultiplets in the
representations
$2\times[(\bf{120},\bf{1})+(\bf{1},\bf{120})]$, $1\times(\bf{16},\bf{16})$,
and 20 singlets. The $U(16)\times U(16)'$ model has an $\omega\rightarrow
1/\omega$ duality which interchanges the two $U(16)$ group factors.
After compactifying on $T^2$ one obtains a $D=4$ model with the tree-level
prepotential given by a special case of eq.(\ref{pp}):
\be
F^{(0)}=SS'U-{\textstyle\frac{1}{2}\sum_i}(SA_i^2+S'A_i^{\prime 2})\ ,
\label{ptree}\ee
where $A$ and $A'$ refer now to $U(16)$ and $U(16)'$ gauge multiplets,
respectively. 
Note that, from the Heterotic point of view, $U(16)'$
has a purely non-perturbative origin. 

In order to determine perturbative corrections to the prepotential in 
Type I theory,
one could in principle follow the method applied on the Heterotic side,
by extracting the one-loop K\"ahler potential $K^{(1)}$
from the universal (gauge group-independent) part of threshold corrections
to gauge couplings \cite{agnt,prep}. In fact,
 the one-loop threshold corrections
have been recently studied in the Coulomb phase of the 
$U(16)\times U(16)'$ model
\cite{bf}. They depend on $U$ and Wilson-line  moduli only,
 which is consistent
with the  general form of perturbative expansion (\ref{fI}). 
Without losing generality, we can focus on the $SU(16)$ subgroup 
originating from N-N boundary conditions.
At zero Wilson lines, the corresponding  gauge coupling takes the form:
\be \frac{4\pi^2}{g^2}= \frac{\pi}{2} S_2+\Delta
\label{g2}\ee
where $S_2$ is the tree-level contribution. The threshold correction 
is\footnote{Here we use the standard field theory normalization
of gauge couplings which amounts
 to multiplying the result of ref.\cite{bf} by a 
factor of 2.}
\be
\Delta=6\int_0^{\infty}\frac{dt}{t}Z(t),\label{del}\ee
where
\be Z(t)=\sum_{p\in \Gamma_2}e^{-\pi t|p|^2/2}
\label{z}\ee
is the partition function of the two-dimensional torus lattice $\Gamma_2$,
with momenta restricted to Kaluza-Klein modes:
\be
p=\frac{m_1+m_2\ubar}{\sqrt{2U_2}\,G^{1/4}}\label{mom}\ee
with integer $m_1$ and $m_2$. Due to this restriction $SL(2,Z)_T$ symmetry
is lost while $SL(2,Z)_U$ remains as a perturbative symmetry.
The integral (\ref{del}) has a logarithmic
infrared divergence at $t\rightarrow\infty$, which reproduces
the correct low-energy running of the gauge coupling with the
beta function coefficient $b=6$.\footnote{The apparent ultraviolet divergence
in eq.(\ref{del}) disappears when the expression is appropriately cut off
\cite{bf}. The potential  divergence is anyway U-independent,
and thus does not affect our discussion here.}

In the Heterotic theory, as mentioned before, the one-loop K\"ahler metric
can be extracted from threshold corrections by using the relation \cite{agnt}
\be
\partial_U\partial_{\ubar}\Delta=-\frac{b}{(U-\ubar)^2}
+4{\pi}^2 K^{(1)}_{U\ubar}\ .\label{rel}
\ee
Using the identity
\be
\partial_U\partial_{\ubar}e^{-\pi t|p|^2/2}
=-\frac{1}{(U-\ubar)^2}\partial_tt^2\partial_te^{-\pi t|p|^2/2}\ ,\label{id}\ee
which follows from eq.(\ref{mom}), we obtain after differentiating
eq.(\ref{del}):
\begin{eqnarray}
\partial_U\partial_{\ubar}\Delta&=&-\frac{6}{(U-\ubar)^2}\int_0^{\infty}
dt\partial_t^2[tZ(t)]\nonumber\\ 
&=&-\frac{6}{(U-\ubar)^2}\partial_t\left.\left(t\sum_{p\in \Gamma_2}
e^{-\pi t|p|^2/2}\right)\right|_0^{\infty}=
-\frac{6}{(U-\ubar)^2}.\label{bound}
\end{eqnarray}
The final result comes from the boundary term at $t=\infty$;
the boundary at $t=0$ does not contribute as one can see easily by
performing a double Poisson resummation in $m_1$ and $m_2$,
cf.\ eq.(\ref{mom}). This result coincides with the
first term of eq.(\ref{rel}). Does this mean that 
$K^{(1)}_{U\ubar}=0$?

The answer turns out to be no. 
In the Heterotic case, the above procedure relied on the fact
that there are no one-loop corrections to the Planck mass \cite{pm}.
In contrast, we will see that such corrections do appear
in Type I theory. As a result, the Type I $S$ field as defined in eq.(\ref{ss})
and below requires a redefinition at the one-loop level in order
to remain an $N=2$ special coordinate. Indeed, assuming that
the Einstein term receives a one-loop correction $\delta$,
so that the coefficient of $R^{(4)}$ in eq.(\ref{bos4}) is
\be
-\frac{1}{2}(e^{-2\phi_4}+\delta)R^{(4)}\ ,
\ee
one has to redefine the dilaton $e^{-\phi_4}\rightarrow e^{-\phi_4}+
\frac{1}{2}\delta e^{\phi_4}$ (to the leading order). 
The gauge coupling $S_2$ of eq.(\ref{ss}) is then redefined as 
$S_2\rightarrow S_2+\sqrt{G}\delta/(2S'_2)$.
As a consequence, the gauge couplings (\ref{g2}) receive
a universal correction which upon using the relation (\ref{rel})
translates to
\be
K^{(1)}_{U\ubar}=\frac{1}{16\pi S'_2}\sqrt{G}\;
\partial_U\partial_{\ubar}\delta\ .\label{final}
\ee

The above equation is also valid in the presence of Wilson lines. 
The momentum lattice $\Gamma_2$ is then  shifted in a way described in
\cite{bf}. Depending on the sector, one has $\Gamma_2\to A_i+\Gamma_2$ or
$A_i+A_j+\Gamma_2$. The $A_i$'s are defined in eq.(\ref{aa}) and the shifted 
lattice
$A_i+\Gamma_2$ is defined with momenta as in eq.(\ref{mom}) with the numerator
replaced by $m_1+m_2\ubar+{\bar A}_i$. It is now easy to verify that 
eq.(\ref{id})
remains valid, hence also eq.(\ref{final}). In the next section we compute the
Planck mass correction $\delta$.

Before concluding this section,
 we would like to make a few comments concerning
the gauge group dependent part of threshold corrections in Type I string
theories. Integrating eq.(\ref{bound}) and using
$SL(2,Z)_U$ symmetry one obtains $\Delta=-b\ln [U_2|\eta(U)|^4]+
\makebox{const}$, where $\eta$ is the Dedekind eta-function. This result is
valid
for any gauge group factor and gives the
$U$-modulus dependence of the one-loop gauge couplings in open
string models, including the case of $S'$-dependent
 tree-level couplings
(\ref{pp}) with $v'\neq 0$. The coefficient $v'$ can be determined by anomaly
cancellation in $D=6$ and was shown \cite{afiq} to be related to the
four-dimensional $N=2$ beta-function, $v'_i-v'_j=(b_i-b_j)/6$. This is
consistent with duality since Type I theory reproduces the familiar result of
Heterotic string models for the group dependent part of threshold corrections
\cite{dkl}, $\Delta_i-\Delta_j=-(b_i-b_j)\ln [U_2T_2|\eta(U)\eta(T)|^4]$,
in the limit $T=S'\to i\infty$.  

\section{One-loop Correction to Planck Mass}

   In order to extract the one-loop correction to Newton's constant,
we consider an amplitude with two external graviton insertions
\be
\partial_U \sum_{\rm one-loop\atop\rm surfaces} \ll
  V_h(p_1,\varepsilon^1)
V_h(p_2,\varepsilon^2) \gg\ 
=\  - {1\over 4}
\varepsilon_{\mu\nu}^1\varepsilon_{\lambda\rho}^2
\eta^{\mu\lambda}p_1^{\rho}p_2^{\nu}\ \partial_U\delta
+{\cal O}(p^4),
\label{amp}
\ee
where $\ll\ \gg$ stands for the path integral over world-sheets
of given topology, $\epsilon^{1,2}$ are the polarization tensors and  
\be
V_h(p,\varepsilon) =  8 \int d^2z\  \varepsilon_{\mu\nu}:
(\dbar x^{\nu}
+{1\over 2} \psibar^{\nu} p\cdot\psibar )
(\partial x^{\mu}-{1\over 2} \psi^{\mu} p\cdot\psi 
)e^{ip\cdot x} : \label{vh}
\ee
is the graviton vertex operator in the zero-ghost picture.
Here, $x^{\mu}$ are the space-time coordinates, 
$\psi^{\mu}$ ($\psibar^{\mu}$) are their left- (right-) moving fermionic
superpartners and $2d^{2}z \equiv dzd\zbar $.
The one-loop surfaces of type-I theory are the torus (${\cal T}$),
annulus (${\cal A}$), M\"obius strip (${\cal M}$)  and Klein
bottle (${\cal K}$). 
Strictly-speaking the amplitude (\ref{amp}) vanishes on shell due to
momentum conservation and the transversality conditions. A correct
procedure is to start  with  the three-point amplitude 
between two gravitons and a U-modulus, which are on-shell  but have
complex momenta. Extracting the desired kinematic structure from
this amplitude gives the same result as the amplitude (\ref{amp}),  if
we  blindly ignore the fact that in this  latter
 $p_1^\mu\varepsilon^2_{\mu\nu}$ should vanish~\cite{agnt}.

In calculating the left-hand side of eq.~(\ref{amp}) one must 
contract at least half of the fermions, or else the  spin-structure
 summation gives  zero. These contractions supply the
desired powers of momenta, so we may set $p=0$ elsewhere to  find
\begin{eqnarray}
\partial_U\delta&=& - 16
\sum_{{\sigma}={\cal A,M,K}}
\int_0^{\infty}\frac{dt}{t}
(2\pi^2t)^{-2}  \partial_U Z(t)
\int d^2z d^2w {1\over 2}
\sum_{s=2,3,4} (-)^s 
{\theta_s^2\over \eta^6}
 Z^{int}_{s,\sigma} \times\nonumber
\\
& &\hspace*{-5mm}\times\left\{
\langle \partial x(z)\partial x(w)\rangle_{\sigma}
\langle\psibar (\zbar )\psibar (\wbar )\rangle^2_{\sigma,\bar s}
 - \langle \partial x(z) \dbar x(\wbar )
\rangle_{\sigma}
\langle\psibar (\zbar )\psi (w)\rangle^2_{\sigma,s} + c.c. \right\}
\label{a1}
\end{eqnarray}
Here
${\theta_s^2/ \eta^6}$ is the oscillator contribution of 
bosonic and fermionic coordinates  of the non-compact space  
plus two-torus; $(-)^s$ is the usual sign of spin-structure summation
which for the desired kinematic structure can be restricted to the
even ones;
 the factor $(2\pi^2t)^{-2}$ comes from the integration
over space-time momenta; 
$Z(t)$ is the sum over torus momenta
which carries all U-dependence and,
in the absence of Wilson lines,  is given by eq.(\ref{z});
finally
$Z^{int}_{s,\sigma}$ is the contribution  of the internal $N=4$ 
superconformal theory describing the $K_3$ compactification to six 
dimensions, including for the annulus 
and M\"obius, the multiplicity of  Chan-Patton states.
 Notice  that we have omitted the torus diagram in the above
expression: this vanishes, as we will argue below, consistently
with the fact that the Einstein term is not renormalized in
$N=2$ heterotic models.

The bosonic  and fermionic
 propagators on ${\cal A,M,K}$ can be obtained from those on the torus
by the method of images \cite{bm}. This is  described in 
detail in the appendix.
Using the fermionic propagators (\ref{pf})
  one can put  the spin-structure summation in  the form
\be
\sum_{s=2,3,4} (-)^s 
{\theta_s^2(0) \over \eta^6}
 Z^{int}_s\times  {1\over 4}
{\theta_s^2(v) {\theta_1^\prime}^2(0)
\over \theta_s^2(0) \theta_1^2(v) } =  \pi^2 Z_{s=1}^{int}
\label{sum}
\ee
We have here used the fact that the partition function of
the internal superconformal theory depends on spin structure only
through the characters of the associated level-one SU(2)
Kac Moody algebra,  so that the entire sum collapses by the
Riemann $\theta$-identity to an index \cite{dkl}. This index is a trace
over open-string Ramond or closed-string Ramond-Ramond states,
weighted with 
the fermion-parity operator $(-)^{F_{int}}$.
It implies that only massless six-dimensional states, that
give rise to N=2 BPS multiplets in four dimensions, contribute 
to the amplitude, as is also the case for threshold corrections
to the gauge couplings  \cite{dl,bf}. Notice also
 the similarity of this result to the analogous expressions for the one 
loop corrections to gauge couplings and K\"ahler metric in
the  heterotic
string  \cite{dkl,agnt,hm}.

To complete the calculation we must still perform the 
 $z$- and $w$-integrals of the  
bosonic correlators of eq.(\ref{a1}).
 The corresponding
calculation for the torus diagram would give zero for the
following reason: fermions can only contract when they are
both holomorphic or antiholomorphic, and 
 $\langle \partial_zx \partial_wx \rangle$ is the derivative
of a periodic function and thus vanishes, when integrated
over the entire torus. This argument does not go through
for the other three one-loop  surfaces, which are 
 obtained by modding out
covering tori with an  appropriate $Z_2$ involution $I_\sigma$. 
This is explained in the appendix, where we also derive
the expressions 
\begin{eqnarray}
\langle \partial x(z)\partial x(w)\rangle_{\sigma}
&=& 
\partial_{z}\partial_{w} P_B(z,w;\tau ) + \frac{\pi}{4\tau_2} \\
\langle \partial x(z)\dbar x(\wbar )\rangle_{\sigma}
&=& 
\partial_{z}\partial_{\wbar} P_B(z,I_{\sigma}(w);\tau )
 - \frac{\pi}{4\tau_2}\ 
\end{eqnarray}
where  $P_B$ is the bosonic propagator on the covering torus, with
modular parameter  
 $\tau = it/2,\,  1/2 + it/2,\,  2it$\  for the surfaces 
${\sigma = \cal A,\, M,\, K}$, respectively. 
Now using the fact that
for  a function $f$ that is periodic on the covering torus
\be
\int_\sigma \partial_w f(w) - \partial_{\bar w}
 f(I_\sigma (w)) = \int_{\cal T} \partial_w f(w) = 0 \ ,
\label{truc}
\ee
we can easily perform the integrals  of the bosonic propagators
in eq. (\ref{a1}) with the
result
\begin{eqnarray}
\int d^2z\int d^2w\  \Bigl\{ 
 \langle \partial x(z)\partial x(w)\rangle_{\sigma}
-\langle \partial x(z)\dbar x(\wbar )\rangle_{\sigma}+c.c. \Bigr\}&=& \\
={\pi\over 4} \tau_2 &=& \cases{&$ \pi t/8$\  {\rm for}\ 
 $\sigma ={\cal A,M}$ \cr
&$\pi t/2$\ {\rm  for}\  $\sigma= {\cal K}$\cr }\nonumber
\end{eqnarray}

Putting together all results, we arrive at our  final expression
for the one-loop renormalization of Newton's constant
\be
\partial_U\delta=-\frac{1}{2\pi}
\Bigl( {1\over 2} Z^{int}_{1,{\cal A}}+
 {1\over 2} Z^{int}_{1,{\cal M}}+ 2 Z^{int}_{1,{\cal K}}
\Bigr)
\int_0^{\infty}\frac{dt}{t^2}\partial_U Z(t)\ ,\label{udel}
\ee
The index discussed previously counts hypermultiplets minus
 the graviton and vector  multiplets in four dimensions \cite{dkl,hm}.
The relative factor of four between surfaces with and without
a boundary, 
accounts for  the fact
that while an open-string hypermultiplet has four Ramond states,
a  closed-string hypermultiplet contains only a single Ramond-Ramond
state \cite{dl,bf}.  
The final result takes thus  the form
\be
\partial_U\delta=-\frac{2}{\pi}
\int_0^{\infty}\frac{dt}{t^2}  \partial_U
\left\{ \sum_{\rm BPS~ hypermultiplets} e^{-\pi tM^2/2}-
\sum_{\rm BPS~ vector~multiplets}e^{-\pi tM^2/2} \right\}\ ,
\label{bps}
\ee
where the masses in this expression originate from  momentum
in the internal two-torus.
This expression is similar to the general formula for the one-loop
$N=2$ prepotential in the Heterotic case \cite{hm}. 

\section{Example of String Triality}

In this section we discuss one specific model which has
simultaneous Type I, Heterotic and Type II descriptions.
On the Type I side it originates from a six-dimensional model
with one tensor multiplet and a completely broken gauge group.
Anomaly cancellation (\ref{ac}) constraints such a model to contain
244 hypermultiplets. This model can be obtained from the
class of orientifold constructions discussed
in section 4, which are based on the $K_3$ orbifold $T^4/Z_2$.
 It belongs to a
subclass which has a perturbative Heterotic description either as
$SO(32)$ or $E_8\times E_8$, compactified on $K_3\times T^2$ with instanton
numbers $(12,12)$ \cite{blpssw}. The 8 five-branes are
 then located half at each
fixed point of the orbifold.
The D-D gauge group is broken down to $U(1)^{16}$, which furthermore
obtain mass by coupling to sixteen hypermultiplets coming from
the closed-string twisted sector \cite{dl}.
 The maximal gauge group in $D=6$ is thus
$SU(16)\times U(1)$ and the hypermultiplet spectrum consists of two
antisymmetric $\bf{120}$'s
from the N-N sector, 16 fundamental $\bf{16}$'s from the
D-N sector and 4 singlets from the closed string sector.
It is easy to see that this 
N-N  gauge group can be broken completely by scalar
vacuum expectation values, leaving exactly 244 hypermultiplets.
Four of those come from the closed string sector,
 while 240 remain from the open strings.

Upon toroidal compactification to $D=4$ one finds also the
3 universal vector multiplets $S,~S'$ and $U$. A Heterotic
-- Type II dual pair with the same massless spectrum has been
considered before in refs.\cite{kv,klm}. On the Type II side,
it corresponds to a IIA compactification on the Calabi-Yau
threefold $WP_{1,1,2,8,12}(24)$ with Hodge numbers $h_{(1,1)}=3$ and  
$h_{(1,2)}=243$.
The first indication that this pair is also equivalent
to the above Type I construction comes from its perturbative
$SL(2,Z)_U$ symmetry as well as from the  $S\leftrightarrow S'$
symmetry which is the remnant of the six-dimensional  
$\omega\rightarrow 1/\omega$ duality. The latter is mapped
to $S\leftrightarrow T$ exchange which was found to be an exact symmetry
of the Calabi-Yau compactification \cite{klm}. In order to
 make a quantitative
comparison, we will first use the formulae derived 
in section 5 to determine the
one-loop correction to the Type I prepotential.

Applying   eq.(\ref{bps}) in the case under consideration,
 one finds 
\be
\partial_U\delta=-\frac{2}{\pi}
\int_0^{\infty}\frac{dt}{t^2}\times  240  \partial_U Z(t)
 ,
\label{bpsex}
\ee
where the torus partition function $Z(t)$ is given in eq.(\ref{z}).
We can now  extract the one-loop K\"ahler metric by  applying
 $\partial_{\ubar}$
to eq.(\ref{udel}) and, using eq.(\ref{bpsex})
 and the identity (\ref{id}), to obtain
\be
\partial_U\partial_{\ubar}\delta=
-\frac{120}{\pi}\frac{1}{U_2^2}\int_0^{\infty}\frac{dt}{t^2}
\partial_tt^2\partial_tZ(t)\ .
\ee
After the change of variables $t=1/l$ and double Poisson resummation
in the $T^2$ partition function (\ref{z}) one finds
\be
\partial_U\partial_{\ubar}\delta=
-\frac{480}{\pi}\frac{\sqrt{G}}{U_2^2}\int_0^{\infty}dl\,l\partial_ll^2
\partial_l{\sum_{n_1,n_2}}'e^{{-4\pi\sqrt{G}\over U_2}|n_1+n_2\ubar|^2l}
=-\frac{240}{\pi^3\sqrt{G}}{\sum_{n_1,n_2}}'
\frac{1}{|n_1+n_2U|^4}
\ee
with the sum running over all integers except for $n_1=n_2=0$.
Substituting this result to eq.(\ref{final})
 we obtain the following one-loop
correction to the K\"ahler metric: 
\be
K^{(1)}_{U\ubar}=-\frac{15}{\pi^4}\frac{1}{S'_2}{\sum_{n_1,n_2}}'
\frac{1}{|n_1U+n_2|^4}\label{gloop}\ .
\ee

We now turn to the Heterotic side of the model. The one-loop
prepotential has been determined in ref.\cite{prep}. In the limit
$T_2\rightarrow\infty$, its third derivative reads
\be
\partial_U^3f_H(U,T_2\rightarrow\infty)=
-\frac{[\partial_Uj(U)]^2}{\pi^2j(U)[j(U)-j(i)]}=4E_4(U)\label{het}\ee
where $j(U)$ is the $SL(2,Z)$ modular function with a simple pole at
infinity while the weight-4 lattice function
\be
E_4(U)=\frac{45}{\pi^4}{\sum_{n_1,n_2}}'\frac{1}{(n_1+n_2U)^4}\ .
\ee
Using the standard $N=2$ formulae one finds that the one-loop corrected
K\"ahler metric is
\be
K_{U\ubar}=-\frac{1}{(U-\ubar)^2}+\frac{1}{S_2}K^{(1)}_{U\ubar}
\label{glooph}
\ee
with $K^{(1)}_{U\ubar}$ given by the same expression as in eq.(\ref{gloop})
with $S'_2$ replaced by $T_2$. We see that the Type I result
(\ref{gloop}) corresponds to the $T\to i\infty$ limit of the Heterotic
case, as expected from duality.

\section{Summary}

In this work, we studied the general features of the effective
field theory describing $N=2$ compactifications of Type I superstrings.
A particular role is played by two dilaton-like fields associated
to continuous Peccei-Quinn symmetries which remain unbroken
in perturbation theory. Under Type I -- Heterotic duality one of
them is mapped to the Heterotic dilaton $S$ and the other 
to the $T$ modulus. 

The one-loop computations presented in sections 5 and 6
provide a strong test of Type I -- Heterotic duality conjecture
for a class of $N=2$ models based on $K_3\times T^2$ compactifications.
Weakly coupled Type I theory is recovered in the weakly coupled regime of 
the Heterotic theory in the limit of
 large K\"ahler modulus $T$ ($T^2$ volume),
provided that the $K_3$ volume of Type I compactification is large
($\omega^4>1)$. When the $K_3$ volume is small,
Type I perturbation theory probes a non-perturbative region
in Heterotic theory.
On the other hand, space-time non-perturbative effects
in Type I theory which are exponentially suppressed at large $T$ 
are mapped to world-sheet instantons on the Heterotic side.
The most interesting conclusion  of this work is the fact that
the type-I prepotential is determined by the renormalization
of Newton's constant and it is related to an index. 
It should be straightforward to extend these
results to other type-I models.
\vskip 1cm

\noindent{\bf Acknowledgements} 

We are grateful to D. L\"ust, S. Ferrara, K.S. Narain and especially 
to A. Sagnotti for very
useful conversations. T.R.T. acknowledges the hospitality of Ecole 
Polytechnique
during the initial stage of this work.
\newpage
\begin{flushleft}
{\large\bf Appendix}\end{flushleft}
\renewcommand{\theequation}{A.\arabic{equation}}
\renewcommand{\thesection}{A.}
\setcounter{equation}{0}

\begin{figure}
\[
\psannotate{\psboxto(0cm;5cm){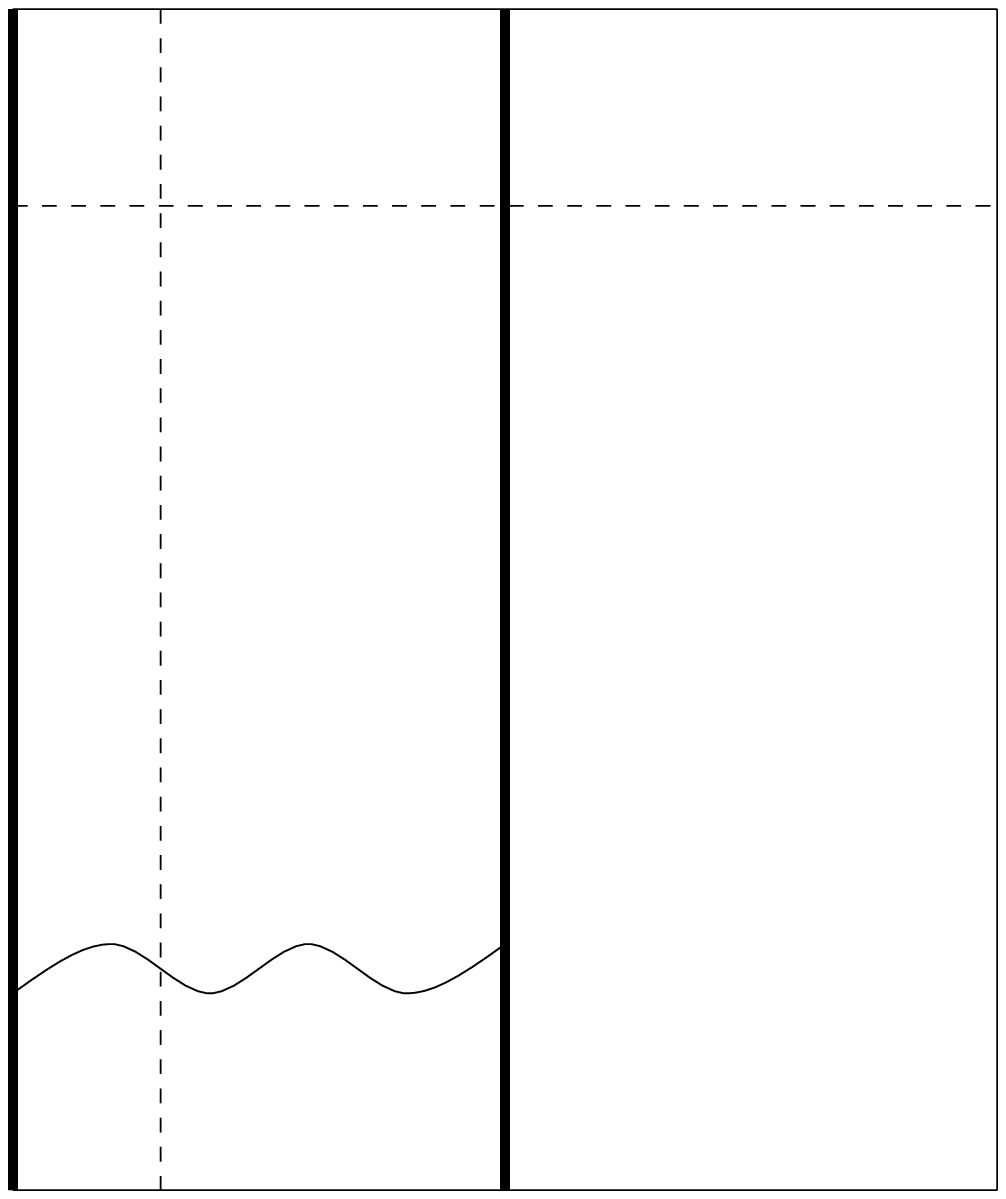}}
{
\at{7.5\pscm}{-2\pscm}{\large Annulus}
\at{6.2\pscm}{11.9\pscm}{$\gg$}
\at{6.2\pscm}{-0.1\pscm}{$\gg$}
\at{7.4\pscm}{6\pscm}{$\bullet$}
\at{9.5\pscm}{6\pscm}{$\bullet$}
\at{6.4\pscm}{5\pscm}{$M$}
\at{8.4\pscm}{5\pscm}{$M'$}
\at{1.8\pscm}{-0.8\pscm}{$0$}
\at{12.2\pscm}{-0.8\pscm}{$1$}
\at{-1.4\pscm}{12\pscm}{$\tau = i\frac{t}{2}$}
\at{8.2\pscm}{9.1\pscm}{$a$}
\at{7.8\pscm}{9.9\pscm}{$>$}
\at{2.8\pscm}{6\pscm}{$b$}
\at{1.6\pscm}{6\pscm}{$\wedge$}
}
\hskip 1.5in
\psannotate{\psboxto(0cm;5cm){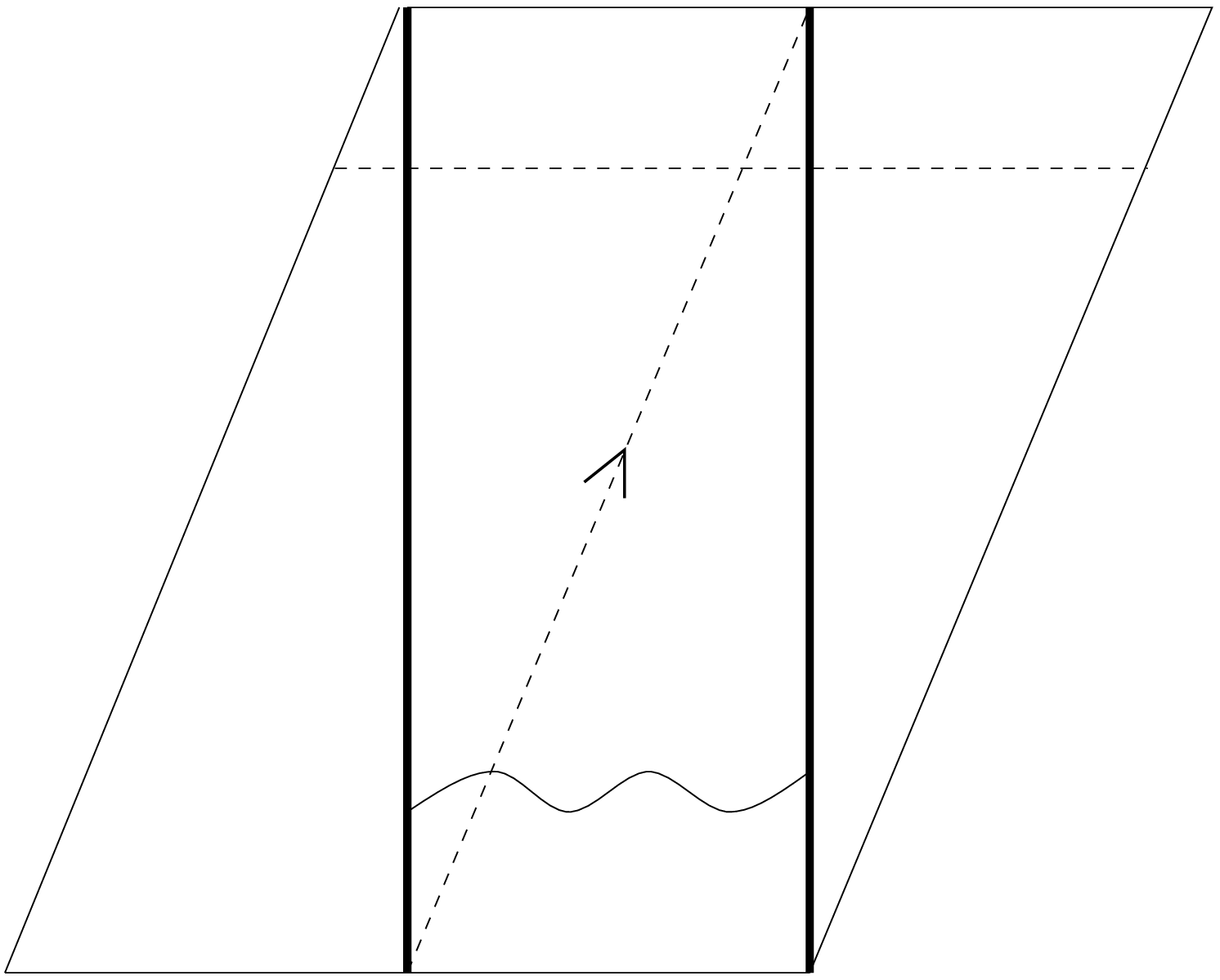}}
{
\at{5.8\pscm}{-2\pscm}{\large M\"obius strip}
\at{10.9\pscm}{11.9\pscm}{$\ll$}
\at{10.7\pscm}{-.1\pscm}{$\gg$}
\at{7\pscm}{7.2\pscm}{$\bullet$}
\at{9.1\pscm}{7.2\pscm}{$\bullet$}
\at{6.0\pscm}{6.2\pscm}{$M$}
\at{8\pscm}{6.2\pscm}{$M'$}
\at{1.3\pscm}{-0.8\pscm}{$0$}
\at{12.1\pscm}{-0.8\pscm}{$1$}
\at{1.1\pscm}{12\pscm}{$\tau = \frac{1}{2}+i\frac{t}{2}$}
\at{12.9\pscm}{9.1\pscm}{$a$}
\at{12.8\pscm}{9.9\pscm}{$>$}
\at{8.4\pscm}{6\pscm}{$b$}
}
\] \\
\[
\psannotate{\psboxto(0cm;5cm){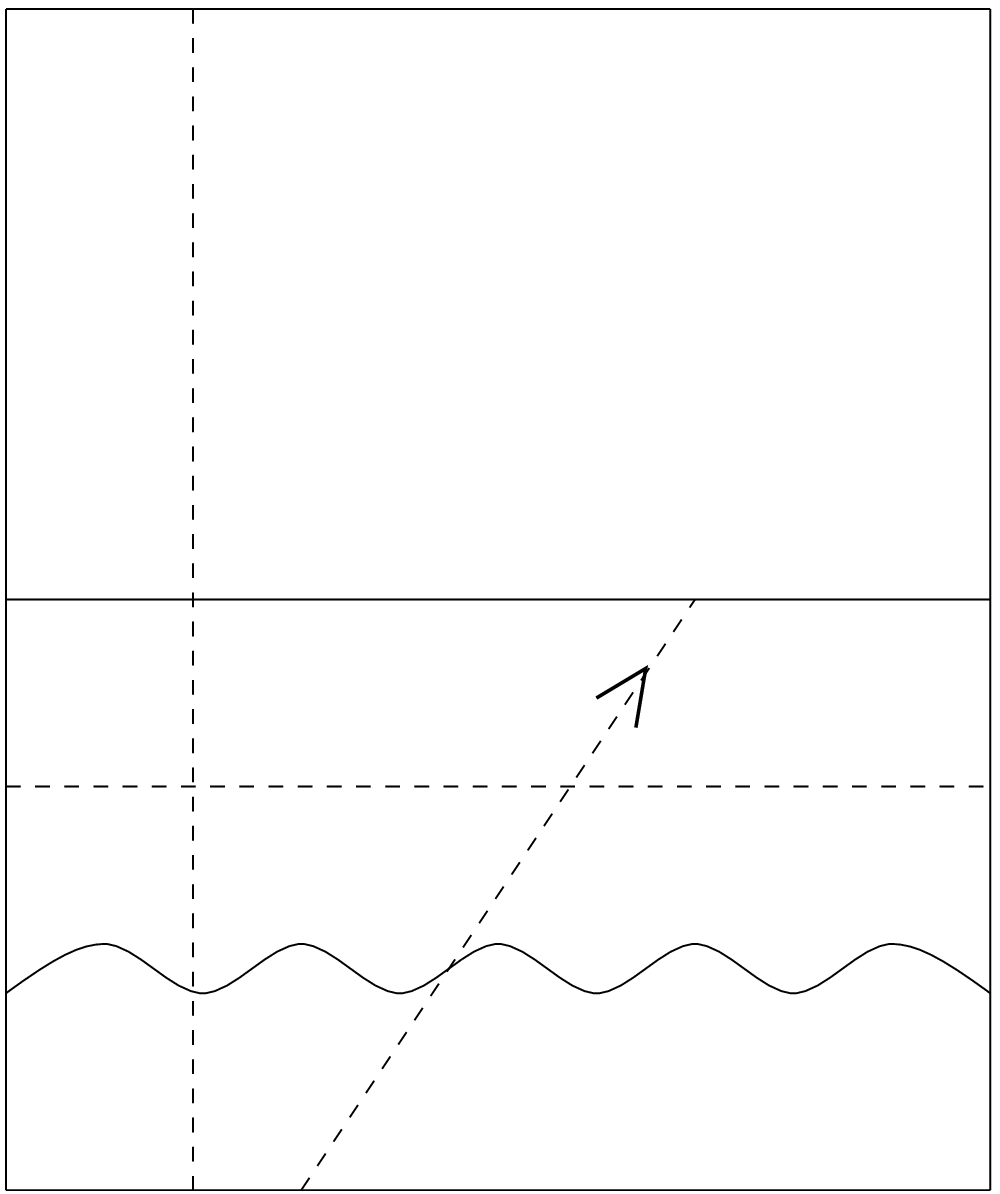}}
{
\at{7.6\pscm}{-2\pscm}{\large Klein bottle}
\at{5.1\pscm}{2.7\pscm}{$\wedge$}
\at{14.7\pscm}{2.7\pscm}{$\wedge$}
\at{9.3\pscm}{5.8\pscm}{$\ll$}
\at{9.1\pscm}{-0.2\pscm}{$\gg$}
\at{4.7\pscm}{5.2\pscm}{$\bullet$}
\at{12.5\pscm}{11.2\pscm}{$\bullet$}
\at{3.7\pscm}{4.4\pscm}{$M$}
\at{11.5\pscm}{10.3\pscm}{$M'$}
\at{2\pscm}{-0.8\pscm}{$0$}
\at{12.5\pscm}{-0.8\pscm}{$1$}
\at{-1.4\pscm}{12\pscm}{$\tau = 2it$}
\at{4.7\pscm}{4.7\pscm}{$a$}
\at{4.3\pscm}{3.9\pscm}{$>$}
\at{3.6\pscm}{8.5\pscm}{$b_{\cal T}$}
\at{2.3\pscm}{8.5\pscm}{$\wedge$}
\at{7.2\pscm}{4.7\pscm}{$b_{\cal K}$}
}
\]

\vskip .3in

\caption{Covering tori and fundamental cells for the
three  one-loop surfaces ${\sigma = \cal A,\, M,\, K}$. 
The cycles are represented by dashed lines.
The points $M'$ are images of $M$ under the appropriate involutions.
The loci of fixed points drawn in thick
 are open-string boundaries.
\label{surfaces}}
\end{figure}

We present here the derivation of various propagators which we use in the
calculation of the one-loop amplitude (\ref{amp}, \ref{a1}) on the annulus
$({\cal A})$, M\"obius strip $({\cal M})$ and
Klein bottle $({\cal K})$. These surfaces can be defined as quotients of tori
under different involutions (see fig.\ref{surfaces})
\be
I_{\cal A}(z)=I_{\cal M}(z)=1-\zbar\ ,\qquad\qquad I_{\cal K}(z)
=1-\zbar+\tau /2\ ,\label{is}
\ee
where $\tau=\tau_1+i\tau_2$ is the modular parameter of the defining torus.
The fundamental cells of the involutions can be  chosen as follows:
$${\cal A}:~~z\in[0,1/2]\times[0,\tau_2] \qquad
{\cal M}:~~z\in [1/2,1]\times[0,\tau_2] \qquad
{\cal K}:~~z\in[0,1]\times[0,\tau_2 /2]\ .$$
Actually in section 5 we use the periodicity properties to make  the
integration region for  the M\"obius strip identical to the one
for the annulus.
The open string boundaries, corresponding  to the loci of fixed
points,  are drawn as thick lines in fig.\ref{surfaces}.
There are no fixed points for   the Klein bottle
representing the evolution and orientation flip of a closed string.
 Notice also  that the
three covering 
 tori are characterized by different modular parameters:  $\tau = it/2,
\,\   1/2 + it/2,\,\   2it$ for the surfaces
 ${\sigma = \cal A,\, M,\, K}$, respectively.

The bosonic correlators can be expressed in terms of the propagator
on the torus $\cal T$
\be 
\langle x(z)x(w)\rangle_{\cal T}=
-\frac{1}{4}
\ln \left|\frac{\theta_1(z-w|\tau)}{\theta'_1(0|\tau)}
\right|^2+ \frac{\pi (z_2-w_2)^2}{2\tau_2}\equiv  P_B(z,w)\ ,
  \label{pbost}
\ee
by symmetrizing  under the corresponding involutions:
\begin{eqnarray}
\langle x(z)x(w)\rangle_{\sigma} &=&\frac{1}{2} \Bigl[ P_B(z,w)+
P_B(z,I_{\sigma}(w))+P_B(I_{\sigma}(z),w)+
P_B(I_{\sigma}(z),I_{\sigma}(w)) \Bigr] \nonumber\\
&=&P_B(z,w)+P_B(z,I_{\sigma}(w))\ .\label{pbos}
\end{eqnarray}
We  follow throughout
the conventions of Green, Schwarz and Witten \cite{gsw}
and we set  $\alpha^\prime = 1/2$.
The above expressions must be supplemented with the usual
holomorphic-regularization prescription, which ensures that
left- and right-movers communicate only through their zero modes.
For the torus this implies that
\be
\langle \partial x(z)\partial x(w)\rangle_{\cal T} =
\partial_{z}\partial_{w} P_B(z,w;\tau) \ ,\ \  
{\rm but}\ \ \ 
\langle \partial x(z)\bar \partial x(w)\rangle_{\cal T}=
 -{\pi\over 4\tau_2}
 \label{pbost1}
\ee
For the other three surfaces we have likewise
\be
\langle \partial x(z)\partial x(w)\rangle_{\sigma} = 
\partial_{z}\partial_{w} P_B(z,w;\tau ) + \frac{\pi}{4\tau_2} =
  -\frac{1}{4}\partial_{z}\partial_{w}
\ln \left|\frac{\theta_1(z-w|\tau)}{\theta'_1(0|\tau)}
\right|^2   + \frac{\pi}{2\tau_2}
\ee
and
\be
\langle \partial x(z)\dbar x(\wbar )\rangle_{\sigma}= 
\partial_{z}\partial_{\wbar} P_B(z,I_{\sigma}(w);\tau ) 
- \frac{\pi}{4\tau_2} =  
 -\frac{1}{4} \partial_{z} \partial_{\wbar}
\ln \left|\frac{\theta_1(z- I_{\sigma}(w)|\tau)}
{\theta'_1(0|\tau)}
\right|^2  - \frac{\pi}{2\tau_2} \ .
\ee
As a check one can  verify that these propagators
have the correct short distance 
singularity
and  periodicity properties on each surface.
Furthermore
the  normal derivatives  on the boundaries vanish, 
consistently with our choice of Neumann boundary conditions
for the non-compact space-time coordinates.

We now turn to fermionic correlators. For
2-dimensional Majorana spinors 
\begin{equation}
\Psi (z,\bar{z})=\left( \begin{array}{c}
                \psi(z) \\
                \tilde{\psi}(\bar{z})
                        \end{array}     \right)
\ee
the propagator on the torus reads
\begin{equation}
\langle\Psi (z,\bar{z})\Psi^T (w,\bar{w})\rangle_{\cal
T} = P_F(s;z,w)\left( \frac{1+\gamma^3}{2} \right) +
\bar{P}_F(\bar{s};\bar{z},\bar{w})
\left( \frac{1-\gamma^3}{2} \right)\ ,
\label{P_F}
\end{equation}
where $\gamma^3=$ diag$(1,-1)$, $s$ and $\bar{s}$ are the even spin
structures of the left and right components, 
\be
P_F(s;z,w)\equiv\langle\psi (z)\psi (w)\rangle^s_{\cal T}
={i\over 2} \frac{\theta_s(z-w|\tau)}{\theta_1(z-w|\tau)}
\frac{\theta'_1(0|\tau)}{\theta_s(0|\tau)}\ .\label{pf}\ee
and
$\theta_s~ (s=2,3,4)$ are the even theta functions.
The propagators on the other  surfaces can be determined again by 
the method of images \cite{bm}. 
The left and right
components of fermions have the same spin structure on all 
covering tori, except for the M\"{o}bius strip for which the
three even spin structures are $(s,\bar{s}) = (2,2),(3,4)$ and $(4,3)$.
One way to understand this subtlety is by noting that for the
M\"{o}bius strip 
$\tau = {1\over 2} + {it\over 2}$ so that
${\bar{\theta}}_3 = \theta_4 $.
The $Z_2$  involutions exchange  left- with right-moving fermions
up to a
subtle choice of signs. One  consistent choice is

\begin{equation}
{\cal I}_\sigma^2 = {\cal I}_\sigma^3  = {\cal I}_\sigma^4
= \left(  \begin{array}{cc}
                                0 & 1 \\
                                1 & 0
                           \end{array} \right) 
\; ,\qquad \sigma={\cal A,\, M}.
\end{equation}
\begin{equation}
{\cal I}_{\cal K}^2 = {\cal I}_{\cal K}^3 =\left(  \begin{array}{cc}
                                0 & -1 \\
                                1 & 0
                           \end{array} \right)\; , \qquad
{\cal I}_{\cal K}^4 = \left(  \begin{array}{cc}
                                0 & 1 \\
                                1 & 0
                           \end{array} \right).
\end{equation}
Symmetrizing the torus propagator under these involutions one finds
\begin{eqnarray}
\langle\psi (z)\psi (w)\rangle_{\sigma}&=&P_F(s;z,w)\nonumber\\
\langle\psi (z)\psibar (\bar{w})\rangle_{\sigma}
&=&P_F(s;z,I_{\sigma}(w))
\label{prop}\\
\langle\psibar (\bar{z})\psibar (\bar{w})\rangle_{\sigma}
&=& {\bar{P}}_F(\bar{s};\bar{z},\bar{w}). \nonumber 
\end{eqnarray}
The reader can check that these 
propagators have the correct pole structure and periodicity properties.

\end{document}